\documentclass[%
 reprint,
 amsmath,amssymb,
 longbibliography,
 aps,
 pre,
]{revtex4-1}
\usepackage{graphicx} 

\usepackage{calc}

\usepackage[normalem]{ulem}
\usepackage{graphicx}
\usepackage{bm}
\usepackage{xcolor}
\usepackage{hyperref}
\usepackage{amsmath}
\usepackage{amssymb}
\usepackage[utf8]{inputenc}
\usepackage{ulem}
\usepackage{xr}
\usepackage{outlines}
\usepackage{bbm}
\usepackage{mathtools}
\usepackage{graphicx}
\usepackage{hhline}
\usepackage{mathrsfs}
\usepackage[normalem]{ulem} 
\usepackage{float} %
\usepackage{tikz}
\usetikzlibrary{arrows}

\def\env{\bm{\xi}}
\def\Z{\mathbb{Z}}

\def\randMeasure{\nu}
\def\expEnv{\mathbb{E}^{\env}}
\newcommand{\expAnnealed}[1]{\mathbb{E}_{\nu}\left[ #1 \right]}

\def\varAnnealed{\mathrm{Var}_{\nu}}

\newcommand{\tiltedExpAnnealedTwo}[1]{\tilde{\mathbf{E}}_{\nu}\left[ #1 \right]}
\def\extCoef{\lambda_\mathrm{ext}}

\def\E{\mathbb{E}}

\def\centeringTerm{c(N)}
\def\driftVar{\varAnnealed\left(\expEnv[Y]\right)}

\newcommand{\diffRandomWalk}[1]{\Delta\left( #1 \right)}
\newcommand{\absMeasure}[1]{\tilde{\mu}\left( #1 \right)}
\newcommand{\deltaIncrement}[1]{\kappa\left( #1 \right)}
\def\annealedMeasure{\mathbf{P}_{\randMeasure}}

\newcommand{\meanOmega}[1]{\bar{\xi}(#1)}
\newcommand{\invMeasure}[1]{\mu({#1})}

\begin{document}

\title{Universal KPZ Fluctuations for Moderate Deviations of Random Walks in Random Environments}
\author{Jacob Hass$^*$, Hindy Drillick$^\dagger$, Ivan Corwin$^\dagger$, Eric Corwin$^*$}
\affiliation{$^*$Department of Physics and Materials Science Institute, University of Oregon, Eugene, Oregon 97403, USA. \\ $^\dagger$Department of Mathematics, Columbia University, New York, New York 10027, USA.}
\date{\today}

\begin{abstract}
    The theory of diffusion seeks to describe the motion of particles in a chaotic environment. Classical theory models individual particles as independent random walkers, effectively forgetting that particles evolve together in the same environment. \emph{Random Walks in a Random Environment} (RWRE) models treat the environment as a random space-time field that biases the motion of particles based on where they are in the environment. We provide a universality result for the moderate deviations of the transition probability of this model over a wide class of choices of random environments. In particular, we show the convergence of moments to those of the multiplicative noise stochastic heat equation (SHE), whose logarithm is the Kardar-Parisi-Zhang (KPZ) equation. The environment only filters into the scaling limit through one parameter, which depends explicitly on the statistical description of the environment. This forms the basis for our introduction, in \cite{hassExtremeDiffusionMeasures2024a}, of the extreme diffusion coefficient. 
\end{abstract}

\maketitle

\section{Introduction} 
Classical diffusion theory is used to describe the statistical behavior of agents in a wide variety of systems with random/chaotic fluctuations such as stock prices \cite{liuAnchoringEffectFirst2017, blackPricingOptionsCorporate1973a}, the movement of photons in a scattering medium \cite{ishimaruDiffusionLightTurbid1989a} and the spread of viruses \cite{kaoVirusDiffusionIsolation2006a}. 
The theory is built upon the assumption that each particle can be modeled as effectively independent random walkers \cite{einsteinZurTheorieBrownschen1906a, einsteinUberMolekularkinetischenTheorie1905a, einsteinTheoretischeBemerkungenUber1907a, langevinTheorieMouvementBrownien1908a, sutherlandViscosityGasesMolecular1893a, sutherlandLXXVDynamicalTheory1905a, vonsmoluchowskiZurKinetischenTheorie1906a, vonsmoluchowskiNotizUiberBerechnung1915a}, thus reducing everything to one parameter---the Einstein diffusion coefficient---that characterizes the variance per unit time of those walkers. Despite the simplicity of this model, it is remarkably effective in describing the statistics of the bulk, or typical diffusing particle in a system of many particles \cite{perrinMouvementBrownienMolecules1910, perrinMouvementBrownienRealite1909, brownXXVIIBriefAccount1828a}.
However, a recent series of works \cite{barraquandRandomWalkBetadistributedRandom2017b, thiery2016exact, ledoussalDiffusionTimeDependentRandom2017c, barraquandModerateDeviationsDiffusion2020c, hassAnomalousFluctuationsExtremes2023, hassFirstpassageTimeManyparticle2024, hassExtremeDiffusionMeasures2024a, dasKPZEquationLimit2023b, parekhInvariancePrincipleKPZ2024a} has provided evidence that the independent random walkers model for many-particle diffusion fails to accurately predict the behavior of extreme particles, i.e., those that travel the furthest or fastest and are often of considerable interest \cite{schussRedundancyPrincipleRole2019a, linnExtremeHittingProbabilities2022a, lawleyDistributionExtremeFirst2020a, lawleyUniversalFormulaExtreme2020a, madridCompetitionSlowFast2020a, basnayakeAsymptoticFormulasExtreme2019a}. Those works have focused on models of \emph{Random Walks in a Random Environment} (RWRE) with environments that are quickly mixing in space and time. Whereas typical particles in such models for many-particle diffusion behave as if the environment was statistically averaged, i.e., reduced to the independent random walker model, the extreme particles manifest novel behavior in the presence of a common random environment. The purpose of this paper is to unwind how the statistical description of the random environment translates into the statistical behavior of the extreme particles.

In the model considered here, an environment is a collection of transition probabilities, indexed by discrete one-dimensional space and time. At each time, particles choose their next spatial location independently according to the transition probabilities at that space-time point. The random environment comes from choosing those transition probability distributions randomly so as to be independent and identically distributed over all of space and time. We consider the many-step transition probability---which is random in light of the random environment. We show that as the time span grows in a scale $N$, the \emph{moderate deviations} of this transition probability in a spatial scale $N^{3/4}$ (i.e., the probability of a single particle moving $N^{3/4}$ to the right of its mean velocity) converges to the solution of the \emph{stochastic heat equation} (SHE), whose logarithm solves the \emph{Kardar-Parisi-Zhang equation}, and which is given by
\begin{equation}\label{eq:SHE}
\partial_T Z = \frac{1}{2} \partial_X^2 Z + \sqrt{2 D_0} Z \eta. 
\end{equation}
Above we have $Z(0,X)=\delta(X)$ (Dirac delta function) initial data, and take $\eta(X,T)$ to be space-time Gaussian white noise (i.e., $\mathbb{E}[\eta(X,T)] = 0$ and $\mathbb{E}[\eta(X,T) \eta(X', T')] = \delta(X-X')\delta(T-T')$). 

The only parameter in this limit is the \emph{noise strength} $D_0 \in \mathbb{R}_{> 0}$, which we determine in Eq. \ref{eq:noiseStrength} explicitly in terms of the statistical description of the random environment. We show convergence at the level of moments, and, though our methods can be applied to general moments, we restrict ourselves to the first and second moments which suffice to pin down the value of $D_0$. Similar results have been demonstrated recently in the mathematics literature in work of Parekh \cite{parekhInvariancePrincipleKPZ2024a}. That work, as well as ours, can be seen as a generalization of the nearest neighbor or sticky-Brownian motion models studied in
\cite{barraquandModerateDeviationsDiffusion2020c, dasKPZEquationLimit2023b, dasKPZEquationLimit2023c} to arbitrary random environments. Since moderate deviations translate into the behavior of the maximum of many draws from a probability distribution, our results translate into results about the statistical behavior of the extreme particles under certain scalings of time and the number of particles \cite{hassExtremeDiffusionMeasures2024a, hassAnomalousFluctuationsExtremes2023, ledoussalDiffusionTimeDependentRandom2017c}, as well as the location of first passage barriers  \cite{hassExtremeDiffusionMeasures2024a, hassFirstpassageTimeManyparticle2024, ledoussalDiffusionTimeDependentRandom2017c} and the correlations between the positions of particles \cite{ledoussalDynamicsEdgeIndependent2024}.

The remainder of the paper proceeds as follows. In Section \ref{sec:RWREModel}, we clearly describe the RWRE model and some relevant notation. Section \ref{sec:MainResults} contains our main results, namely the convergence of the first two moments and the determination of $D_0$. Our approach is explained in Section \ref{sec:Overview}---in particular we use a variant of the replica method to relate moments to discrete local times, and then employ two probabilistic tools---the Tanaka formula for Brownian local times, and the stationary measures for certain Markov chains---to relate this to the replica formulas for the SHE moments. 
Section \ref{sec:secondMoment} fleshes out this local time convergence approach, and Section \ref{sec:TailProbability} relates those calculations back to the moments of the moderate deviations.

\section{The RWRE Model}\label{sec:RWREModel}
The random environment in which we will consider random walks is defined as $\env\coloneqq\{\xi_{t,x}: x \in \mathbb{Z}, t \in  \mathbb{Z}_{\geq 0} \}$ where each $\xi_{t,x}$ is a probability distribution on $\mathbb{Z}$. The $\xi_{t,x}$ are themselves random, chosen to be independent and identically distributed over all choices of $t$ and $x$, with a distribution $\nu$ that completely determines the RWRE model. Given an environment $\env$, we will consider $N$ independent random walkers; when a walker is at position $x$ and $t$ time, they choose how far to jump in the next time interval according to the distribution $\xi_{t,x}$. Different walkers at the same site use the same distribution $\xi_{t,x}$ but sample from it independently. Thus, there are two levels of randomness in the model, that of the environment in which all particles evolve together and that of the independently sampled walker trajectories. We introduce some notation to keep track of these two levels of randomness and illustrate it with an example.

We let $\mathcal{M}_1( \Z)$ denote the space of probability distributions on $\Z$ and $\randMeasure$ a probability distribution on $\mathcal{M}_1(\Z)$. Each choice of $\randMeasure$ corresponds to a different law on the random environment $\env$, where we sample $\xi_{t,x} \in \mathcal{M}_1(\Z)$ according to $\randMeasure$, independently for each $(t,x)$. We will restrict ourselves to only consider $\nu$ which are supported on finite range probability distributions, i.e., if  $\xi$ is distributed according to $\randMeasure$ then for some $M$ sufficiently large, $\xi(j) = 0$ for all $j > M$. We let $\expAnnealed{\bullet}$ denote the expectation of a function $\bullet$ of $\env$ with respect to the product measure where each $\xi_{t,x}$ is independent with distribution $\nu$. To illustrate this notation, let us note one example. For integer $k>0$, and $\alpha_{-k},\ldots, \alpha_{k}>0$, let $\big(\xi(-k),\ldots,\xi(k)\big)$ be Dirichlet distributed with the specified $\alpha$ parameters  (and let all other $\xi(j)=0$ for $j\notin \{-k,k\}$). Precisely, restricted to the simplex where $\xi(-k),\ldots,\xi(k)\in [0,1]$ and $\xi(-k)+\cdots+\xi(k)=1$, the probability density function is proportional to $\prod_{j=-k}^{k}\xi(j)^{\alpha_j-1}$. This defines a random probability distribution on $\Z$ (or rather, $\{-k,\ldots, k\}$) and hence defines one choice of $\randMeasure$.

Given an instance of the environment $\env$, we will consider $N \in \mathbb{N}$ random walkers evolving independently with jump distributions determined by $\env$, and denote the corresponding probability measure by $\mathbb{P}^{\env}$. More precisely, $\mathbb{P}^{\env}$ is a probability measure on the sample space $(\mathbb Z^N)^{\mathbb Z_{\geq 0}}$, where we think of an element  $\mathbf R = (R^1,\ldots,R^N)$ of this space as the time-indexed spatial trajectory of $N$ walkers $R^1(t),\ldots, R^{N}(t)$ for $t\in \mathbb{Z}_{\geq 0}$.
For a given realization $\env$ of the environment, we define $\mathbb{P}^{\env}$ to be the measure on $(\mathbb Z^N)^{\mathbb Z_{\geq 0}}$ such that $(R^1, \ldots, R^N)$ are distributed as $N$ independent random walks all started at $0$, with independent increments given, for each $k\in \{1,\ldots, N\}$ by 
\begin{equation*}
\mathbb{P}^{\env}(R^k(t+1)=x+i \mid R^k(t) = x) = \xi_{x,t}(i).
\end{equation*}
In other words, each walker uses the jump distribution at their current time and position to determine the size of their next jump. When the environment $\env$ is random, the measure $\mathbb{P}^{\env}$ is random as well. We will be interested below in understanding the random probability distribution of a single walker, i.e., $\mathbb{P}^{\env}(R^k(t)=x)$. In that case, we will adopt the notation $R(t)=R^1(t)$, dropping the superscript. 

We define the \emph{annealed} probability measure  $\annealedMeasure$ on the same $N$-path sample space  $(\mathbb Z^N)^{\mathbb Z_{\ge 0}}$ by averaging $\mathbb{P}^{\env}$ over $\env$, according to the earlier described $\randMeasure$-dependent product measure $\mathbb{E}_{\randMeasure}$. 
In other words, we define the annealed probability measure $\annealedMeasure (\cdot) := \expAnnealed{\mathbb{P}^{\env}(\cdot)}$ and the corresponding expectation $\mathbf E_{\nu}[\cdot] = \expAnnealed{\mathbb{E}^{\env}[\cdot]}$. For any $k \leq N$, we call  the process $ (R^1, \ldots, R^k)$ with law given by $\annealedMeasure$ the \emph{k-point motion}.

We denote the \emph{average} environment as $\bar{\xi}=\expAnnealed{\xi_{t,x}}$, which can also be thought of as the ensemble-averaged environment, meaning what you see if you average over a large swath of space and time. This average is the same at each site, hence, no $t,x$ subscript is needed. For simplicity, we will only consider models with $\nu$ that have a drift-free average environment, i.e., such that $\sum_{i \in \Z} \meanOmega{i} i = 0$, though we expect similar results to hold for general $\nu$ after changing to a suitable moving reference frame. 

Since we will make extensive use of them, we describe here the one- and two-point motions $R^1$ and $(R^1,R^2)$ under the annealed measure $\annealedMeasure$. The one-point motion $R^1$ is an independent and identically distributed (i.i.d.) increment random walk that jumps from position $x$ to position $x+j$ with probability $\meanOmega{j}$ for all $j\in \mathbb{Z}$. The same is true marginally of $R^2$, however, it is not quite true that $R^1$ and $R^2$ are independent. When  $R^1(t)\neq R^2(t)$, $(R^1,R^2)$ evolves according to independent $\bar{\xi}$ distributed increments, i.e., if $R^1(t)=x$ and $R^2(t)=y$ then $R^1(t+1)=x+j$ and $R^2(t+1) = y + k$ with probability given by the product $\bar{\xi}(j) \bar{\xi}(k)$.  However, when $R^1(t)=R^2(t)=x$, the probability that $R^1(t+1)=x+j$ and $R^2(t+1)=x+k$ is equal to $\expAnnealed{\xi_{t,x}(j) \xi_{t,x}(k)}$.

It follows from the above two-point motion transition formulas that we can think of them as a pair of \emph{sticky} random walks since when the two walkers are at the same site at time $t$, they are more likely to also be at the same site at time $t+1$ than two independent random walks. An important object in our analysis will be the difference $V(t) := R^1(t) - R^2(t)$ between the two-point motion $(R^1,R^2)$. In particular, we will study the gap $\Delta(t):= |V(t)|$. The walk $V(t)$ is a Markov chain on $\mathbb{Z}$ where the transition probability from state $i$ to $j$ is
\begin{equation}\label{eq:jump_rates}
    p(i,j) = \begin{cases}
        \sum_{k \in \Z}\expAnnealed{\xi_{t,x}(k) \xi_{t,x}(k-j)}
 &\text{if $i=0$} \\
  \sum_{k \in \Z}\bar{\xi}(k) \bar{\xi}(k+ j-i)  &\text{if $i \neq 0$},  \end{cases}
\end{equation}
which only depends on $j-i$ and whether or not $i = 0$. Furthermore, for $i, j \neq 0$, we have  $p(i,j)=p(j,i)$.

\section{Main result: RWRE limit to SHE}\label{sec:MainResults}
Based on previous results for nearest neighbor RWRE jump models \cite{thiery2016exact,barraquandModerateDeviationsDiffusion2020c, ledoussalDiffusionTimeDependentRandom2017c, dasKPZEquationLimit2023b}, we expect that in the moderate deviation regime, the tail probability will have environmental dependent fluctuations that have a SHE scaling limit. We will precisely state the scaling of this tail probability and demonstrate convergence to the SHE at the level of the first and second moments. Our methods can be readily generalized to show convergence of higher moments, but the first and second moments suffice in identifying the apriori unknown noise strength coefficient $D_0$.  

We study the (random) tail probability (recalling our shorthand convention that $R(t)=R^1(t)$)
\begin{equation}\label{eq:tailProb}
    \mathbb{P}^{\env}\left( R(NT) \geq N^{3/4}T + \frac{\sum_{i\in Z} \meanOmega{i} i^3}{2 (2D)^2} N^{1/2} T  + \sqrt{2DN} X\right)
\end{equation}
in the limit that $N \rightarrow \infty$, where $N \in \Z_{>0}$, $T \in N^{-1} \Z_{\geq 0}$, $D \coloneqq \frac{1}{2} \sum_{i \in \Z} \meanOmega{i} i^2$ is the diffusion coefficient of a random walk in the average environment, and $X \in \mathbb{R}$ such that the right-hand side of Eq. \ref{eq:tailProb} is an integer. We have dropped the superscript on the random walk in Eq. \ref{eq:tailProb} since all $N$ random walks are i.i.d. We find that the fluctuations of Eq. \ref{eq:tailProb} converge to the SHE in Eq. \ref{eq:SHE} with noise strength 
\begin{equation}\label{eq:noiseStrength}
D_0 = \frac{\extCoef}{(2 D)^{3/2}}
\end{equation}
where $\extCoef$ is a characteristic length defined as follows. Let $Y$ be a random variable distributed according to $\xi_{t,x}$, which can be thought of as a single step of a random walk $R^1$. Further let  $\invMeasure{l}$ be the unique invariant measure of $V(t)$ with the normalization $\mu(0) = 1$ (we address the necessary modifications for the case where the invariant measure of $V(t)$ is not unique in Section \ref{sec:InvariantMeasure}). We then have
\begin{equation}\label{eq:lambdaDef}
    \extCoef = \frac{\driftVar}{\sum_{l=0}^{\infty} \absMeasure{l}\mathbf{E}_\nu[\diffRandomWalk{t+1} - \diffRandomWalk{t} \mid \diffRandomWalk{t} = l]},
\end{equation}
where
\begin{align*}
    \driftVar &= \expAnnealed{\left( \sum_{i \in \Z} \xi_{t,x}(i) i \right)^2}, \\ 
    \absMeasure{l} &=  \begin{cases} 
        1 & \text{if $l=0$} \\
       2 \mu(l) &  \text{if $l >0$}.
    \end{cases}
\end{align*}
Although the form of $\extCoef$ is rather complex, it simplifies significantly for a number of distributions, which we discuss in \cite{hassExtremeDiffusionMeasures2024a}. 

The term $\driftVar$ satisfies $\driftVar \in [0, 2 D]$ and represents the variance, over all environments, of the drift of a single jump of a random walk. When $\driftVar = 2D$, the jump distribution is only supported on a single site, so the walks are perfectly sticky. In the limit that $\driftVar \rightarrow 2 D$, sticky Brownian motion can be recovered as studied in \cite{dasKPZEquationLimit2023c}. When $\driftVar = 0$, the drift of the system is deterministic. Since we only study environments that are net drift free, this means $\expEnv[Y] = 0$ with probability $1$ \cite{parekhInvariancePrincipleKPZ2024a, hassUniversalKPZFluctuations2024}. 

The invariant measure, $\absMeasure{l}$, can be interpreted as how much time two particles spend a distance $l$ apart as compared to being at the same site in the long time limit. Experimentally, this could be measured as follows. Start two particles in the same environment. Let them diffuse for a long time and then measure their distance apart. Repeat this many times in different environments to build a histogram of the distance apart. After normalizing the histogram to $1$ when the particles are at the same location, the histogram represents the invariant measure, $\absMeasure{l}$. Alternatively, the invariant measure could be measured by observing a large system of diffusing particles. Letting the system run for a long time and then building a histogram of the distances between pairs of particles yields the invariant measure, $\absMeasure{l}$.

The term $\mathbf{E}_{\nu}[\diffRandomWalk{t+1} - \diffRandomWalk{t} | \diffRandomWalk{t} = l]$ does not depend on $t$ as we are conditioning on $\diffRandomWalk{t} = l$. This term quantifies the change in the distance between two random walks.  Since we are assuming a finite size jump distribution, the expected value will be 0 for large enough $l$, and thus, the sum in the denominator will be finite. 

Our results agree with those for nearest neighbor RWRE models \cite{barraquandModerateDeviationsDiffusion2020c, dasKPZEquationLimit2023b}. Our results also agree with and are a specific case of a more general class of RWRE models studied in \cite{parekhInvariancePrincipleKPZ2024a}.

\section{Overview of Our Derivation}\label{sec:Overview}

To show convergence of the tail probability to the SHE, we work with a more general setup of which the tail probability in Eq. \ref{eq:tailProb} is a special case. We study the probability mass function smoothed by a spatial test function. We include this smoothing because the probability mass distribution at a single lattice site is too noisy to work with (i.e., it depends on the order one behavior of the noise). The tail probability can be recovered by choosing the spatial test function to sum over all lattice sites in the tail, as discussed below. Of course, any physical measurement of diffusion involves some smoothing, so this is a natural lens through which to study the model.

Since the RWRE probability mass function is given in terms of a discrete path integral through the random environment, its moments admit representations in terms of interacting random walks, where the interaction relates to the law of the environment and is active when the walks coincide. This description is a discrete analog of the replica method formulas for moments of the SHE in terms of Brownian motions interacting through their local times. The argument presented below identifies a scaling under which the discrete moment formulas converge to their continuum SHE counterparts. It turns out that convergence arises only in a specific \emph{moderate deviation} regime in the tail of the RWRE probability distribution.

To start, we re-express the location about which we are studying the tail probability as a window of size  $\sqrt{2DN}$ about $x = c(N) T$, where 
$$c(N) = N^{3/4} + \frac{\sum_{i \in \Z} \meanOmega{i} i^3}{2 (2D)^2} N^{1/2}.$$ 
This specific choice places us in the moderate deviation regime (the central limit regime is $x$ of order $N^{1/2}$, and the large deviation regime is $x$ of order $N$) and is justified in detail below. The probability mass distribution at this location in the tail decays to first order like a Gaussian as $N\rightarrow\infty$, mimicking the behavior of the tail in the central limit regime. About this Gaussian behavior there are random fluctuations in the probability mass function. To study these fluctuations, we rescale the probability mass function by the first-order Gaussian behavior so the fluctuations stay of order one. This asymptotic Gaussian behavior is encoded in the prefactor $C(N, T, X)$, whose particular functional form is chosen for future mathematical convenience 

Specifically, we study 
\begin{align}\
    &\mathscr{U}_N(T, \phi) := \label{eq:UdefExp}\\ 
    \nonumber&\expEnv\left[C\left(N, T, \frac{R(NT)- \centeringTerm T}{\sqrt{2DN}}\right) \phi\left( \frac{R(NT ) -  \centeringTerm T }{\sqrt{2DN}}\right)\right],
\end{align}
where $\phi: \mathbb{R} \rightarrow \mathbb{R}$ is a spatial test function (i.e. a smooth and compactly supported function), $R=R^1$ is a single random walk in the random environment $\env$
and \begin{equation*}
C(N, T, X) \coloneqq \frac{\mathrm{exp}\left\{\displaystyle\frac{\centeringTerm}{2D N^{1/4}} T + \frac{1}{\sqrt{2D}} N ^{1/4} X\right\}}{\left(\displaystyle\sum_{i\in\Z} \meanOmega{i} \displaystyle \mathrm{exp}\left\{ \frac{i}{2DN^{1/4}}\right\} \right)^{NT}}.
\end{equation*}
If one thinks of $\phi$ as bounded support bump functions, then $\mathscr{U}_N(T, \phi)$ is a weighted (by the $C(N,T,X)$ factor) probe of the distribution of $R(NT)$ in the vicinity of $c(N)T$ and in the scale $\sqrt{2 D N}$.

We now give our general strategy to show the moments (for different $k$) $\expAnnealed{\mathscr{U}_N(T, \phi)^k}$ of Eq. \ref{eq:UdefExp} converge to those of the SHE. We fully realize this for the first and second moments, though the approach works similarly for general moments. Consider the first moment. It is immediate, see \ref{eq:ExpU}, that $\expAnnealed{\mathscr{U}_N(T, \phi)}$ can be expressed in terms of an expectation over the one-point motion. Then, due to its multiplicative nature, the $C(N,T,X)$ prefactor in the definition of $\mathscr{U}_N(T, \phi)$ can be absorbed as a tilt of the jump distribution of the one-point motion, and hence in the limit yields a Brownian expectation.

Before going to the $k=2$ case, let us note that the  $k$-th moment of the SHE integrated against a spatial test function can be written (via the replica method) in terms of the expectation of the local time of $k$ Brownian motions
\begin{align} \label{eq:SHEmoments}
&\mathbb{E}\left[\left(\int_{\mathbb{R}}\phi(X)Z(X,T)dX\right)^k\right] = \\
\nonumber &\mathbf{E}\left[\Phi(\vec{B})\exp\left\{D_0 \sum_{i<j} L^{B^i-B^j}(T) \right\}\right]
\end{align}
where $\mathbb{E}$ on the left is the expectation over the noise $\eta$ of the SHE, and on the right $\Phi(\vec{x}) := \phi(x_1)\cdots\phi(x_k)$, and the expectation $\mathbf{E}$ is over independent Brownian motions $B^1,\ldots B^k$ with $L^{B^i - B^j}(t)$ their pair local time at zero, defined as follows. The local time of a space-time continuous Brownian motion $B(t)$ with variance $\sigma^2 t$ is 
\begin{equation}\label{eq:localtime}
    L^B(t) \coloneqq \lim_{\epsilon\rightarrow 0^+} \frac{\sigma^2}{2 \epsilon} \int_0^t \mathbbm{1}_{\left\{-\epsilon < B(s) < \epsilon \right\}} ds. 
\end{equation}

For $k=2$ (and higher), the moment is rewritten in terms of an expectation with respect to the $2$-point motion. The same tilting argument works provided the two-point motions occupy different spatial locations. When they are at the same site, the two-point motion jump distribution (i.e., $p(i,j)$ from \ref{eq:jump_rates} with $i=0$) has some residual effect after tilting which can be written, as in Eq. \ref{eq:discreteSecondMoment}, in terms of a local time contribution.  The tilted two-point motion clearly converges to independent Brownian motions, so the whole challenge is to understand how the discrete local time converges to a limiting Brownian local time. 

To illustrate this challenge, note that a random walk that lives entirely on the odd sublattice of $\mathbb{Z}$ will have zero discrete local time at 0, yet will still converge to a Brownian motion. More relevant to the current situation, the introduction of some stickiness at zero for a random walk will not impact its Brownian limit (provided the stickiness is not tuned in the scaling limit), but will cause the discrete local time at 0 to converge to a constant multiple of the Brownian local time at 0. That constant dilation factor of the local time depends on the degree of stickiness.
 
Thus, to address the discrete to continuous local time convergence---in particular computing the dilation factor (which translates into the noise strength coefficient $D_0$)---we develop an argument based on a discrete version of the \emph{Tanaka formula} and the \emph{Doob-Meyer decomposition}, both tools based on studying random walks and Brownian motions as \emph{martingales}. This will identify a discrete local time, not entirely concentrated at 0, which converges to the Brownian local time at 0. Then, we will use the invariant measure of the gap between the two-point motion to identify what portion of that discrete local time comes from the discrete local time at 0 (which is what arises in our moment formulas). This will yield the desired dilation factor and explains the form of $D_0$ given earlier. The details of this argument are presented below.

\begin{widetext}
\section{Convergence of the First Moment} \label{sec:firstMoment}
The convergence of the first moment will show that our choice of $C(N,T,X)$ is the correct prefactor to ensure $\mathscr{U}_N(T, \phi)$ converges to the first moment of the SHE. We begin by averaging $\mathscr{U}_N(T, \phi)$ in Eq. \ref{eq:UdefExp} over the random environment to obtain
\begin{equation}\label{eq:ExpU}
\expAnnealed{\mathscr{U}_N(T, \phi)} = \mathbf E_{\nu} \left[C\left(N, T, \frac{R(NT)- \centeringTerm T}{\sqrt{2DN}}\right)\phi\left( \frac{R( NT ) -  \centeringTerm T }{\sqrt{2DN}}\right)\right].
\end{equation}
The notation used above should be recalled from Section \ref{sec:RWREModel}.
In particular, $R=R^1$ is the one-point motion under the annealed measure $\mathbf{P}_\randMeasure$ and is the expectation with respect to that measure $\mathbf{E}_\randMeasure$.

We now absorb the prefactor into the expectation by interpreting it as an \emph{exponential tilting} of the i.i.d. increments of the one-point motion. We start by breaking up the one-point into its increments, $R(NT) = \sum_{i=1}^{NT} Y_i$ where $Y_i$ are independent and identically distributed according to $\bar{\xi}$. Thus, we can rewrite
\begin{align*}
C\left(N, T, \frac{R(NT)- \centeringTerm T}{\sqrt{2DN}}\right) &= \frac{\exp\left\{ \displaystyle\frac{R( NT )}{2D N^{1/4}}\right\}}{\left(\displaystyle\sum_{i\in\Z} \meanOmega{i} \mathrm{exp}\left\{\frac{i}{2DN^{1/4}}\right\}\right)^{NT}} =\prod_{i=1}^{NT} \frac{\displaystyle \exp\left\{ \displaystyle\frac{Y_i}{2D N^{1/4}}\right\}}{\displaystyle\sum_{i\in\Z} \meanOmega{i} \mathrm{exp}\left\{\frac{i}{2DN^{1/4}}\right\}}.
\end{align*}
Since the $Y_i$ are independent and identically distributed according to $\bar{\xi}$, this factor can be absorbed as a tilt of the jump distribution for $Y_i$. Define the tilted measure $\tilde{\mathbf{E}}_\randMeasure$ under which the $Y_i$ are independent and identically distributed but now with the probability that $Y_i=j$ for $j\in \mathbb{Z}$ given by
\begin{equation}\label{eq:tiltedjump}
    \frac{\displaystyle \meanOmega{j} \exp\left\{\displaystyle\frac{j}{2D N^{1/4}}\right\}}{\displaystyle\sum_{i\in\Z} \meanOmega{i} \mathrm{exp}\left\{\frac{i}{2DN^{1/4}}\right\}}.
\end{equation}
Notice how the prefactor was chosen to make sure this tilting results in a probability measure
Recalling that $R(NT) = \sum_{i=1}^{NT} Y_i$, we thus have shown that Eq. \ref{eq:ExpU} can be rewritten in terms of the tilted measure as
\begin{equation}\label{eq:firstMomentTestFunction}
    \expAnnealed{\mathscr{U}_N(T, \phi)} = \tilde{\mathbf{E}}_{\nu} \left[\phi\left( \frac{R( NT ) -  \centeringTerm T }{\sqrt{2DN}}\right)\right].
\end{equation}

Under the tilted measure, the centered and scaled one-point motion converges to a Brownian motion, i.e., 
\begin{equation}\label{eq:Brownianconv}
    \lim_{N\to \infty} \frac{R(NT) - \centeringTerm T}{\sqrt{2DN}} = B(T)
\end{equation}
where $B$ is a standard unit variance Brownian motion. To see this, since the increments of $R$ are independent and identically distributed, it suffices to check that that $\frac{R(NT)- \centeringTerm T}{\sqrt{2DN}}$ has mean $0$ and variance $T$, at least up to terms that vanish as $N \rightarrow \infty$. Under the tilted distribution Eq. \ref{eq:tiltedjump}, the increments of $R(NT)$ have mean 
\begin{align*}
    \frac{\sum_{i \in \Z}\meanOmega{i} i \exp\{\frac{i}{2D N^{1/4}}\}}{\sum_{j\in\Z} \meanOmega{j} \exp\{\frac{j}{2D N^{1/4}}\}} &= \frac{\sum_{i \in \Z} \meanOmega{i} i \left( 1 + \frac{i}{2D N^{1/4}} + \frac{i^2}{4 D N^{1/2}} + \mathcal{O}\left( N^{-3/4} \right)\right)}{\sum_{j\in\Z} \meanOmega{j} \left( 1 + \mathcal{O}\left( N^{-1/2}\right)\right)} \\
    &= \sum_{i \in \Z} \meanOmega{i} \frac{i^2}{2DN^{1/4}} + \sum_{i \in \Z} \meanOmega{i} \frac{i^3}{4 D^2 N^{1/2}} + \mathcal{O}\left(N^{-3/4}\right) \\
    &= \frac{\centeringTerm}{N} + \mathcal{O}\left( N^{-3/4}\right)
\end{align*}
where $\mathcal{O}(x)$ denotes terms of order $x$ and lower. 
The second moment of the increments of $R(NT)$ are
\begin{align*}
    \frac{\sum_{i \in \Z}\meanOmega{i} i^2 \exp\{\frac{i}{2D N^{1/4}}\}}{\sum_{j\in\Z} \meanOmega{j} \exp\{\frac{j}{2D N^{1/4}}\}}&= \frac{\sum_{i \in \Z} \meanOmega{i} i^2 \left( 1 + \mathcal{O}\left( N^{-1/4} \right)\right)}{\sum_{j\in\Z} \meanOmega{j} \left(1 + \mathcal{O}\left( N^{-1/2}\right)\right)} \\
    &= 2D + \mathcal{O}\left( N^{-1/4}\right).
\end{align*}
Thus, under the tilted measure $R(NT)$ has mean $\centeringTerm T + \mathcal{O}(N^{1/4})$ and variance $2DNT + \mathcal{O}(N^{3/4})$, or equivalently $\frac{R(NT)- \centeringTerm T}{\sqrt{2DN}}$ has a mean of order $\mathcal{O}(N^{-1/4})$ and variance $T + \mathcal{O}(N^{-1/4})$. 

Combining Eq. \ref{eq:Brownianconv} with Eq. \ref{eq:firstMomentTestFunction}, we see the first equality \begin{equation}
    \lim_{N \to \infty}\expAnnealed{\mathscr{U}_N(T, \phi)} =  \mathbf E \left[ \phi(B(T))\right] = \mathbb{E}\left[\int_{\mathbb{R}}\phi(X)Z(X,T)dX\right] .
\end{equation}
Here $\mathbf E$ is the expectation with respect to a standard Brownian motion staring at $0$, and the second equality (to the the first moment of the SHE) follows from in Eq.\ref{eq:SHEmoments}.

\section{Convergence of the Second Moment}\label{sec:secondMoment}
We now show that the second moment of $\mathscr{U}(T, \phi)$ converges to the second moment of the SHE with the strength of the noise given by $D_0 = \frac{\extCoef}{(2D)^{3/2}}$. In doing so, we identify the characteristic length scale $\extCoef$.

Recall that given an instance of the environment $\env$, the random walks $R^1, R^2, \ldots$ are independent and identically distributed. Thus, in Eq. \ref{eq:UdefExp}, we could have just as well replaced $R$ by $R^1$ or $R^2$ without changing anything. Doing that, and owing to the independent (given $\env$) of $R^1$ and $R^2$, it follows that 
\begin{align*}
\mathscr{U}_N(T, \phi)^2 &= \mathbb{E}^{\env}\left[C\left(N, T, \frac{R^1( NT ) -  \centeringTerm T}{\sqrt{2DN}}\right) C\left(N, T, \frac{R^2( NT ) -  \centeringTerm T}{\sqrt{2DN}}\right) \phi\left(\frac{R^1( NT ) -   \centeringTerm T }{ \sqrt{2DN}} \right) \phi\left(\frac{R^2( NT ) -  \centeringTerm T }{ \sqrt{2DN}} \right)\right] .
\end{align*}
This is the first step of the \emph{replica method}. 

The next step is to take the expectation over the random environment to get a formula for the second moment. Using the explicit form of the prefactor $C$, this yields
\begin{align}\label{eq:beforeMSub}
\expAnnealed{\mathscr{U}_N(T, \phi)^2} &=\mathbf{E}_{\nu}\left[ \frac{\mathrm{exp}\left\{\frac{R^1( NT ) + R^2( NT )}{2D N^{1/4}}\right\}}{\left(\sum_{i \in \Z} \meanOmega{i} \mathrm{exp}\left\{\frac{i}{2D N^{1/4}} \right\}\right)^{2NT}} \phi\left(\frac{R^1( NT ) -  \centeringTerm T }{ \sqrt{2DN}} \right) \phi\left(\frac{R^2( NT ) -  \centeringTerm T }{ \sqrt{2DN}} \right)\right],
\end{align}
where the pair $(R^1, R^2)$ is the two-point motion defined earlier. As we did for the first moment, we want to absorb the prefactor as a tilt of the transition probabilities for the two-point motion. When $R^1\neq R^2$, this works exactly as before since the two random walks take independent jumps. When $R^1=R^2$, the two jumps are no longer independent, and an additional factor is needed to ensure that the tilting results in a probability measure. This produces a discrete local time at 0 term in our moment formula, see Eq. \ref{eq:secondform} below. 

To derive Eq. \ref{eq:secondform}, first observe that the denominator inside the expectation in Eq. \ref{eq:beforeMSub} can be broken into $NT$ equal factors with each term given by 
\begin{equation}
    \left(\sum_{i \in \Z} \meanOmega{i} \mathrm{exp}\left\{\frac{i}{2D N^{1/4}} \right\}\right)^2 = \sum_{i, j \in \Z} \meanOmega{i} \meanOmega{j} \mathrm{exp}\left\{\frac{i + j}{2D N^{1/4}} \right\}
\end{equation}
after expanding into a double sum. When $R^1(t) \neq R^2(t)$, this is the right normalization for the tilting factor needed to produce a probability measure for the increments of $R^1$ and $R^2$. When $R^1(t) = R^2(t)$, we require a different normalization to get a tilted probability measure since $\meanOmega{i} \meanOmega{j}$ should be replaced by $\expAnnealed{\xi_{t,x}(i) \xi_{t,x}(j)}$ (as the walks are at the same site).

To account for this, we really should have started with a different prefactor
\begin{equation}\label{eq:secondMomentLocalTimeFactor}
   C\left(N, T, \frac{R^1( NT ) -  \centeringTerm T}{\sqrt{2DN}}\right) C\left(N, T, \frac{R^2( NT ) -  \centeringTerm T}{\sqrt{2DN}}\right)  \exp\left\{-g\left((2D)^{-1} N^{-1/4}\right)\sum_{i=0}^{ NT  -1}\mathbbm{1}_{\{R^1(i) = R^2(i) \}}\right\}
\end{equation}
where 
\begin{equation}\label{eq:gDef}
g(\lambda) = \log\left( \sum_{i, j\in\Z} \expAnnealed{\xi_{t,x}(i) \xi_{t,x}(j)} e^{\lambda(i+j)} \right) - 2 \log\left( \sum_{i \in \Z} \bar{\xi}(i) e^{\lambda i}\right).
\end{equation}

If we use this tilting factor, then we obtain a tilted version of the two-point motion which is a Markov chain with the following transition probabilities: When $R^1(t)  \neq R^2(t)$, then $R^1(t+1) = R^1(t)+i$ and $R^2(t + 1) = R^2(t)+ j$ with probability
\begin{equation}\label{eq:tiltedMeasure1}
    \frac{\meanOmega{i}\meanOmega{j}\mathrm{exp}\left\{\frac{i + j}{2D N^{1/4}} \right\}}{\sum_{k, l \in \Z} \meanOmega{k} \meanOmega{l} \mathrm{exp}\left\{\frac{k + l}{2D N^{1/4}} \right\}}.
\end{equation}
When $R^1(t) = R^2(t) = x$, then $R^1(t+1) = x + i$ and $R^2(t+1) = x+ j$ with probability
\begin{equation}\label{eq:tiltedMeasure2}
\frac{\expAnnealed{\xi_{t,x}(i) \xi_{t,x}(j)}\mathrm{exp}\left\{\frac{i + j}{2D N^{1/4}} \right\}}{\sum_{k, l \in \Z} \expAnnealed{\xi_{t,x}(k) \xi_{t,x}(l)} \mathrm{exp}\left\{\frac{k + l}{2D N^{1/4}} \right\}}.
\end{equation}

Noting the overload with the notation from the first moment calculation, we let  $\tiltedExpAnnealedTwo{\bullet}$ denote the expectation value with respect to the tilted probability distribution on $(R^1,R^2)$ given by the above transition probability. 

Taking into account the $g$ factor we included in Eq. \ref{eq:secondMomentLocalTimeFactor}, we find that
\begin{align}\label{eq:secondform}
&\expAnnealed{\mathscr{U}_N(T, \phi)^2} =\\
\nonumber&\qquad\tiltedExpAnnealedTwo{\exp\left\{g\left((2D)^{-1} N^{-1/4}\right)\sum_{i=0}^{ NT  -1}\mathbbm{1}_{\{R^1(i) = R^2(i) \}}\right\} \phi\left(\frac{R^1( NT ) -  \centeringTerm T }{ \sqrt{2DN}} \right) \phi\left(\frac{R^2( NT ) -  \centeringTerm T }{ \sqrt{2DN}} \right)}.
\end{align}
By Taylor expansion we find that $g((2D)^{-1} N^{-1/4}) = \frac{\expAnnealed{\left( \sum_{i \in \Z} i \xi_{t,x}(i)\right)^2}}{(2D)^2 N^{1/2}} + \mathcal{O}(N^{-3/4}) = \frac{\driftVar}{(2D)^2 N^{1/2}} + \mathcal{O}(N^{-3/4})$ where the second equality comes from observing that $\driftVar = \expAnnealed{\left( \sum_{i \in \Z} i \xi_{t,x}(i)\right)^2}$. The fact that this term behaves like $N^{-1/2}$ is key since the discrete local time (i.e., $\sum_{i=0}^{ NT  -1}\mathbbm{1}_{\{R^1(i) = R^2(i) \}}$) needs to be scaled by that factor to have a limit. This consideration is what forces the scaling regime for SHE convergence to be in the $N^{3/4}$-depth moderate deviations of the tail probability.

Substituting this expansion in Eq. \ref{eq:secondform} and using $\approx$ to denote that we have dropped the lower order terms, we have
\begin{equation}\label{eq:discreteSecondMoment}
\expAnnealed{\mathscr{U}_N(T, \phi)^2} \approx \tiltedExpAnnealedTwo{\exp\left\{\frac{\driftVar}{(2D)^2 N^{1/2}} \sum_{i=0}^{ NT  -1}\mathbbm{1}_{\{R^1(i) = R^2(i) \}}\right\} \phi\left(\frac{R^1( NT ) -  \centeringTerm T }{ \sqrt{2DN}} \right) \phi\left(\frac{R^2( NT ) -  \centeringTerm T }{ \sqrt{2DN}} \right)}.
\end{equation}
Notice that Eq. \ref{eq:discreteSecondMoment} looks like a discrete analog of the second moment of the SHE in Eq. \ref{eq:SHEmoments} since $\sum_{i=0}^{ NT  -1}\mathbbm{1}_{\{R^1(i) = R^2(i) \}}$ is the discrete local time of $R^1(i) - R^2(i)$ at 0. Furthermore, besides their sticky interaction when $R^1=R^2$, the tilted two-point motion behaves like independent random walks just as in the first moment case. The only way that the sticky interaction could impact the scaling limit is if it is scaled to become stronger as $N\to\infty$ (i.e., so that the time they stay together increases in the scaling limit), or if the size of the jump from $R^1=R^2$ were scaled to become longer as $N\to\infty$ (i.e., so as to result in a discontinuous jump in the limit process). This is not the case, and hence under the measure $\tilde{\mathbf{E}}_{\nu}$, the pair $\left(\frac{R^1(NT) - c(N)T}{\sqrt{2DN}}, \frac{R^2(NT) - c(N)T}{\sqrt{2DN}}\right)$ converge to independent variance one Brownian motions, as in the first moment case.

Our final step is to identity the scaling limit of the discrete local time $\displaystyle \sum_{i=0}^{ NT  -1}\mathbbm{1}_{\{R^1(i) = R^2(i) \}}$, namely that it converges to a constant (which we identity) times the local time of the difference of two independent standard Brownian motions. This is the most subtle and interesting part of the argument since it more broadly explains in what sense discrete local times converge to their continuum Brownian analogs. In particular, although the discrete random walks converge to independent Brownian motions, the discrete local time need not converge to the local time of their Brownian motion limits. For example, consider a simple symmetric random walk on $\mathbb{Z}$ modified so as to stay at 0 with probability $1/2$ and go to $\pm 1$ each with probability $1/4$ (i.e., it is sticky at the origin). The discrete random walks will converge to Brownian motion, but the discrete local time will converge to a constant (explicitly calculable and exceeding one) times the local time of standard Brownian motion. Identifying this constant which rescales the local time limit is the key to identifying the noise strength of the SHE. Specifically, in what follows, we show that 
    \begin{equation}\label{eq:localTimeConvergence}
        \lim_{N \to \infty} \frac{1}{\sqrt{2DN}} \sum_{i=0}^{ NT  -1} \mathbbm{1}_{\{R^1(i) = R^2(i)\}} = \frac{1}{\sum_{l=0}^{\infty} \absMeasure{l} \mathbf E_{\nu} \left[\diffRandomWalk{t+1} - \diffRandomWalk{t} \mid \diffRandomWalk{t} = l\right]} L_0^{B^1- B^2}(T). 
    \end{equation}
    where $B^1$ and $B^2$ are independent standard Brownian motions starting at $0$. Notice that the prefactor of the local time on the right-hand side is the denominator of $\extCoef$ in Eq. \ref{eq:lambdaDef}. 
    We show this in Section \ref{sec:localTime}.

Putting the above together, we conclude that
\begin{equation*}
\lim_{N\to \infty} \expAnnealed{\mathscr{U}_N(T, \phi)^2} = \mathbf{E}\left[e^{\frac{\extCoef}{(2D)^{3/2}} L_0^{B^1 - B^2}(T)} \phi(B^1(T)) \phi(B^2(T))\right]
\end{equation*}
where $\mathbf{E}$ is the expectation with respect to two independent Brownian motions $B^1$ and $B^2$ start at $0$ and with variance $1$; and $\extCoef$ is given by Eq. \ref{eq:lambdaDef}. This is indeed the second moment of the multiplicative stochastic heat equation defined in Eq. \ref{eq:SHEmoments} with noise strength given by Eq. \ref{eq:noiseStrength}. 

A similar argument applies for the case of general moments. A priori, when dealing with higher moments, there are terms that come from multi-particle interactions, e.g., when $R^1=R^2=R^3$. Each of these contributions needs to be accounted for when writing down the tilted measure and contribute differently to the local time factor. However, in our scaling, these different factors end up factorizing as $N\to\infty$ and thus only contribute in the form of two-body local times, as needed to recover the general moment formula Eq. \ref{eq:SHEmoments}.

\subsection{Convergence to Local Time}\label{sec:localTime}
The purpose of this section is to demonstrate the claimed convergence of Eq. \ref{eq:localTimeConvergence}.
Recall $V(t) = R^2(t) - R^1(t)$, so $\sum_{i=0}^{ NT  -1} \mathbbm{1}_{\{R^1(i) = R^2(i)\}} =  \sum_{i=0}^{ NT  -1} \mathbbm{1}_{\{|V(i)| = 0\}}$ is the sum of occupation times at $0$ for the random walk $V$. To derive Eq. \ref{eq:localTimeConvergence}, we identify a discrete analog of \emph{Tanaka's formula} to identify a smoothed, discrete local time that converges to the local time of Brownian motion. We then use the invariant measure of the two-point gap process, $\diffRandomWalk{t} = |V(t)|$, to relate the smoothed, discrete local time to the discrete local time at $0$.  

Before giving Tanaka's formula, we recall the definition of a martingale. A continuous time martingale is a time-parameterized collection of random variables, $(Y_t)_{t\geq 0}$, which obeys the martingale property whereby $\mathbb{E}\left[Y_{t} \mid \mathcal{F}_s \right] = Y_s$ for all $t>s\geq 0$, where $\mathcal{F}_s$ is the $\sigma$-algebra generated by $Y_r$ for all $s\geq r\geq 0$. In words, this property says that the expected value at time $t$, given the history of the process up to time $s$, is the value at time $s$. Brownian motion $B_t$ with drift zero is an example of a martingale, as is $B_t^2 - t$ (the quadratic martingale) or $e^{\lambda B_t - \lambda^2 t/2}$ (the exponential martingale) for any $\lambda$.

Tanaka's formula gives a decomposition of the absolute value of a Brownian motion into the sum of a martingale and Brownian local time. More precisely, given a Brownian motion $B(t)$, the Tanaka formula states that
\begin{equation}\label{eq:Tanaka}
    |B(t)| = \int_0^t \text{sgn}(B_s) dB_S + L^B(t),
\end{equation}
where $L^{B}(t)$ is the local time at zero, defined in Eq. \ref{eq:localtime}, and 
$$\text{sgn}(x) = \begin{cases}  +1 & x  > 0 \\ 0 & x= 0 \\ -1 & x < 0\end{cases}.$$ 
The stochastic integral $\int_0^t \text{sgn}(B_s) dB_S$ is a martingale, and, in fact, is another Brownian motion. 

Tanaka's formula gives a way to decompose $|B(t)|$ into the sum of a martingale (the stochastic integral) and an increasing \emph{predictable process} (the local time). Such a decomposition is called the \emph{Doob-Meyer decomposition}. A key fact is that the Doob-Meyer decomposition of a \emph{submartingale} (a process such as $Y_t = |B(t)|$ that satisfies $\mathbb{E}\left[Y_{t} \mid \mathcal{F}_s \right] \geq  Y_s$ for all $t>s\geq 0$) is unique under fairly general assumptions. Therefore, if we can obtain \emph{any} decomposition of $|B(t)|$ into the sum of a martingale term and an increasing predictable process, we can identify that increasing predictable process as Brownian local time.

Motivated by this fact, we obtain a Doob decomposition for $\Delta(t) = |V(t)|$ by identifying a discrete analogue of Tanaka's formula. This decomposition involves the sum of a martingale and a discrete local time term. We then take the term-by-term limit of this decomposition. We know that $\Delta(t)$ will converge to $|B(t)|$, where $B(t)$ will be a Brownian motion with variance $2$. The martingale term will converge to a limiting martingale, and the limit of the discrete local time term will still be an increasing predictable process. Hence, by the uniqueness of the Doob-Meyer decomposition for $|B(t)|$, we will conclude that the limit of the discrete local time term is Brownian local time.  

There are several mathematical subtleties that we brush over but quickly note here. From Eq. \ref{eq:discreteSecondMoment}, we are concerned with convergence of an expectation with respect to the titled measure $\tilde{\mathbf{E}}_\nu$ of an expression involving the expectation of the local time at zero for $\Delta(t)$. The argument presented below shows that this local time under the untilted measure $\mathbf{E}_\nu$ converges in distribution to a constant multiple of Brownian local time. The replacement of the tilted measure by its untilted counterpart is mostly out of convenience (the argument could be done under the tilted measure too) and is justified since as $N\to \infty$, the tilted ($N$-dependent) measure converges to the untitled one. This can be seen by expanding $e^x$ around $x=0$ to simplify the exponential in the tilted measure. Then the sums in the denominators of \eqref{eq:tiltedMeasure1} and \eqref{eq:tiltedMeasure2} evaluate to 1, and we recover the transition probabilities of the untilted two-point motion. The issue around convergence of expectations of exponentials would require some further careful justification that we do not pursue here.

Turning to the details of this argument, we decompose $\Delta(t)$ as follows: 
\begin{equation}\label{eq:discreteTanaka}
    \Delta(t)=M(t) + \sum_{i=0}^{t-1}\sum_{l =0}^\infty \deltaIncrement{l}\mathbbm{1}_{\{\Delta(i)=l \}} 
\end{equation}
where we define 
$$
\kappa(l) = \mathbf{E}_\nu\left[\Delta(i+1) - \Delta(i) \big| \Delta(i) = l\right]\qquad \textrm{and}\qquad M(t):= \Delta(t) - \sum_{i=0}^{t-1}\sum_{l =0}^\infty \deltaIncrement{l}\mathbbm{1}_{\{\Delta(i)=l \}}.
$$ 
Note that $\Delta(i+1) - \Delta(i)$ does not actually depend on $i$ since we are conditioning on $\Delta(i) = l$. 

To see that $M(t)$ is a martingale with respect to the untilted measure $\mathbf{E}_\nu$, we rewrite $M(t)$ as
\begin{align*}
   M(t) &= \Delta(t)- \sum_{i=0}^{t-1}\sum_{l =0}^\infty\mathbf{E}_\nu\left[\Delta(i+1) - \Delta(i) \mid  \Delta(i) = l\right]\mathbbm{1}_{\{\Delta(i)=l \}}  \\
    &= \Delta(t) - \sum_{i=0}^{t-1}\mathbf{E}_\nu\left[\Delta(i+1) - \Delta(i) \mid \Delta(i) \right]\\
\end{align*}
where we have taken the sum over $l$ so that we now condition on the random value $\diffRandomWalk{i}$. Notice $\Delta(i+1) - \Delta(i)$ is independent of the values of $\Delta(1), \ldots, \Delta(i-1)$ and only depends on $\Delta(i)$. Thus, we can rewrite   
\begin{align*}
    M(t) &=\Delta(t) - \sum_{i=0}^{t-1}\mathbf{E}_\nu\left[\Delta(i+1) - \Delta(i) \mid  \Delta(1), \ldots, \Delta(i) \right].
\end{align*}
It follows from this and telescoping that $M(t)$ is a martingale. 

We now show that the term $\sum_{i=0}^{t-1}\sum_{l =0}^\infty \deltaIncrement{l}\mathbbm{1}_{\{\Delta(i)=l \}}$ is increasing in $t$ and predictable. We first show that $\kappa(l)\geq 0$. To see this, we use our results from \cite{hassExtremeDiffusionMeasures2024a} which show $\kappa(l)$ simplifies to
\begin{equation}
\kappa(l) = \begin{dcases} 
      \sum_{i, j \in \Z}|i-j|\expAnnealed{\xi(i) \xi(j)} & l=0 \\
      \sum_{|i-j| > l} (|i-j| - l) \meanOmega{i}\meanOmega{j} & l>0 \\
   \end{dcases}.
\end{equation}
Thus, $\kappa(l) \geq 0$ and hence $\sum_{i=0}^{t-1}\sum_{l =0}^\infty \deltaIncrement{l}\mathbbm{1}_{\{\Delta(i)=l \}}$ is a sum of non-negative terms which increases in time. Furthermore, this process is predictable as it only depends on the values of $\Delta(0) \ldots \Delta(t-1)$. Therefore, \eqref{eq:discreteTanaka} is a decomposition of $\Delta(t)$ into the sum of a martingale and an increasing predictable process.

We look at the limiting behavior of this decomposition under a diffusive scaling and match it term by term with Tanaka's formula in Eq. \ref{eq:Tanaka}. As argued in the paragraph after Eq. \ref{eq:discreteSecondMoment}, the term $\Delta(t)=|R^1(t)-R^2(t)|$ converges under diffusive scaling to $|B^1(t) - B^2(t)|$. Since the term $M(t)$ is a martingale, it should likewise converge to a limiting martingale. Therefore, the second term on the right-hand side of Eq. \ref{eq:discreteTanaka} should converge to the local time $L^{B^1 - B^2}(t)$ in Eq. \ref{eq:Tanaka} (where we take $B=B^1-B^2$). Thus, taking $t = NT$, under the diffusive scaling
\begin{align}\label{eq:localtimeLimit1}
    \lim_{N \to \infty}  \frac{1}{\sqrt{2DN}}\sum_{l =0}^\infty \deltaIncrement{l} \sum_{i=0}^{NT-1}\mathbbm{1}_{\{\Delta(i)=l \}} = L^{B^1 - B^2}(T).
\end{align}
This completes the first step towards establishing Eq. \ref{eq:localTimeConvergence}. Of course, that result calls for taking the limit of the discrete local time at $0$, 
\begin{equation}\label{eq:localtimeLimit2}
    \lim_{N \to \infty} \frac{1}{ \sqrt{2DN}} \sum_{i=0}^{Nt-1} \mathbbm{1}_{\{\Delta(i) = 0\}},
\end{equation}
whereas Eq. \ref{eq:localtimeLimit1} is in terms of a combination of local time at every position $l \in \Z_{\geq 0}$. 

To compare the limits in \eqref{eq:localtimeLimit1} and \eqref{eq:localtimeLimit2}, we will show in Section \ref{sec:InvariantMeasure} that for $l \geq 0$, 
\begin{equation}\label{eq:ratiolimit}
\lim_{t \to \infty}\frac{\sum_{i=0}^{t} \mathbbm{1}_{\{\Delta(i) = l\}}}{\sum_{i=0}^{t} \mathbbm{1}_{\{\Delta(i)  = 0\}}} = \tilde{\mu}(l),
\end{equation}
where $\tilde{\mu}$ is the normalized invariant measure of $V(t)$, see the discussion after Eq. \ref{eq:lambdaDef}.
For large $t$, we can therefore approximately relate the local time at $l$ and at $0$ so that
\begin{equation}\label{eq:approx}
    \sum_{i=0}^{t}\mathbbm{1}_{\{\Delta(i)  = l\}} \approx \tilde \mu(l) \sum_{i=0}^{t}\mathbbm{1}_{\{\Delta(i)  = 0\}}.
\end{equation}
Using this approximation, we obtain
\begin{equation*}
    \sum_{l =0}^\infty \deltaIncrement{l}  \sum_{i=0}^{t-1} \mathbbm{1}_{\{\Delta(i) =l \}} \approx \left(\sum_{l=0}^{\infty}\deltaIncrement{l}\absMeasure{l}\right)\sum_{i=0}^{t-1} \mathbbm{1}_{\{\Delta(i) = 0 \}}
\end{equation*}
Substituting this into Eq. \ref{eq:localtimeLimit1} and rearranging, we obtain
\begin{equation}
    \lim_{N \to \infty} \frac{1}{ \sqrt{2DN}} \sum_{i=0}^{Nt-1} \mathbbm{1}_{\{\Delta(i) = 0\}} = \frac{1}{\sum_{l=0}^{\infty}\deltaIncrement{l}\absMeasure{l}} L^{B^1 - B^2}(t).
\end{equation}
After substituting in our definitions of $\kappa(l)$ and $\diffRandomWalk{t}$, we conclude that 
\begin{equation*}
\lim_{N \to \infty} \frac{1}{\sqrt{2DN}} \sum_{i=0}^{Nt-1} \mathbbm{1}_{\{R^1(i) = R^2(i)\}} =  \frac{1}{\sum_{l=0}^{\infty} \absMeasure{l} \mathbf E_{\nu} \left[\diffRandomWalk{t+1} - \diffRandomWalk{t} \mid \diffRandomWalk{t} = l\right]} L^{B^1 - B^2}(t).
\end{equation*}
which indeed matches Eq. \ref{eq:localTimeConvergence}.

\subsubsection{Invariant Measure of V}\label{sec:InvariantMeasure}
In this section, we demonstrate the claim in Eq. \ref{eq:ratiolimit}.
If a Markov chain is irreducible (starting from any state, the chain can eventually reach any other state)  and recurrent (it returns to every state infinitely many times), then it has a unique invariant measure up to a constant multiple. As explained earlier, we consider $V$ under the untilted measure $\mathbf{E}_\nu$ on the two-point motion rather than the tilted probability distribution. We can see that $V(t)$ is recurrent as follows: When $V(t)$ is away from $0$ its increments are i.i.d. and have mean $0$ and thus, by the Chung-Fuchs theorem \cite{chungDistributionValuesSums1951}, the walk will almost surely eventually return to $0$. Once at $0$, it will eventually leave, and the above argument can be repeated to show that $V(t)$ will return to $0$ infinitely many times, which suffices to show recurrence. 
However, $V(t)$ may not always be irreducible. For example, consider the case where $R^1(t)$ and $R^2(t)$ only take nearest neighbor jumps. Then $V(t)$ can only take jumps that are multiples of $2$, so if it starts from $0$, it can only visit sites in the sublattice $2\Z$.

Let us first deal with the case where $V(t)$ is irreducible so that it has a unique (up to a constant multiple) invariant measure. Let
$$T_0 = 0, \quad  T_k = \inf\{n > T_{k-1}: V(n) = 0\} $$ so that $T_k$ is the $k$th return time to $0$. Define $Y_k^l = \sum_{i=T_k}^{T_{k+1}-1}\mathbbm{1}_{\{V(i) = l\}}.$ For an irreducible and recurrent Markov chain, we know that the measure $\mu$ on $\mathbb Z$ defined by $\invMeasure{l} := \E[Y_0^l]$ is an invariant measure for $V$ where $\E$ is the expectation of the walker starting at $0$. Furthermore, since $V$ is a Markov chain, the random variables $(Y_k^l)_{k \geq 0}$ are i.i.d. for a fixed $l$. 

We have that 
\begin{equation}\label{eq:returnTime}
\frac{Y_0^l + \cdots + Y_{n-1}^l}{n} = \frac{\sum_{i=0}^{T_{n}-1}\mathbbm{1}_{\{V(i) = l\}}}{\sum_{i=0}^{T_{n}-1}\mathbbm{1}_{\{V(i) = 0\}}}.
\end{equation}
The equality in the denominator is due to the fact that we visit $0$ exactly $n$ times before time $T_n-1$. By the law of large numbers for i.i.d. random variables, 
\begin{equation}\label{eq:InvMeasure}
\lim_{n \to \infty}\frac{Y_0^l + \cdots + Y_n^l}{n} = \E[Y_0^l] =\invMeasure{l}. 
\end{equation}
almost surely. On the other hand, we have that 
\begin{equation}\label{eq:ratioLocalTimeAndReturnTime}
    \lim_{n \to \infty}\frac{\sum_{i=0}^{T_n - 1}\mathbbm{1}_{\{V(i) = l\}}}{\sum_{i=0}^{T_n - 1}\mathbbm{1}_{\{V(i) = 0\}}}=  \lim_{t \to \infty}\frac{\sum_{i=0}^{t}\mathbbm{1}_{\{V(i) = l\}}}{\sum_{i=0}^{t}\mathbbm{1}_{\{V(i) = 0\}}}.
\end{equation}
Thus, by combining Eqs. \ref{eq:returnTime}, \ref{eq:InvMeasure}, and \ref{eq:ratioLocalTimeAndReturnTime}, we obtain 
\begin{equation}\label{eq:localTimeInvMeasure}
    \lim_{t \to \infty}\frac{\sum_{i=0}^{t}\mathbbm{1}_{\{V(i) = l\}}}{\sum_{i=0}^{t}\mathbbm{1}_{\{V(i) = 0\}}} = \mu(l).
\end{equation}

Note that by symmetry $\mu(l) = \mu(-l)$. Furthermore, for $l= 0$, we have 
$$
\mathbbm{1}_{\{\Delta(i) = l\}} = \mathbbm{1}_{\{V(i) = l\}}
$$ 
and for $l > 0$,
$$\mathbbm{1}_{\{\Delta(i) = l\}} = \mathbbm{1}_{\{V(i) = l\}}+\mathbbm{1}_{\{V(i) = -l\}}.$$
Putting this all together with Eq. \ref{eq:localTimeInvMeasure}, we conclude that 
\begin{align*}
\lim_{t \to \infty}\frac{\sum_{i=0}^{t} \mathbbm{1}_{\{\Delta(i) = l\}}}{\sum_{i=0}^{t} \mathbbm{1}_{\{\Delta(i) = 0\}}} & =  \tilde{\mu}(l)
\end{align*}
where 
\begin{align*}
    \absMeasure{l} &=  \begin{cases} 
        1 & \text{if $l=0$} \\
       2 \mu(l) &  \text{if $l >0$}.
    \end{cases}
\end{align*}

Finally, we consider the case where $V(t)$ is not irreducible. We can decompose $\Z$ into the union of finitely many closed and irreducible sets  (sublattices) $E_i$.  Let $E_0$ be the set containing $0$. Since we have $V(0) = 0$, we know that $V(t) \in E_0$ for all $t$. Let $\mu$ be the unique invariant measure of the Markov chain $V(t)$ when restricted to the state space $E_0$. All of the above analysis goes through, but the sum $\sum_{l=0}^{\infty} \kappa(l)\tilde{\mu}(l)$ gets replaced by $\sum_{l \in E_0}\kappa(l)\tilde{\mu}(l)$. An example of this situation is when $R^1$ and $R^2$ are nearest neighbor random walks. In this case, $V(t)$ is restricted to the even integer sublattice, i.e., $E_0 = 2\mathbb Z$.

\section{Convergence of the Tail Probability to the SHE}\label{sec:TailProbability}
The above discussion shows that the first two moments of the rescaled probability mass distribution converge to the moments of the SHE. In this section, we discuss the convergence of the tail probability, which can be written as 
\begin{equation}\label{eq:tailphi}
    \mathbb{P}^{\env}( R(NT) \geq \centeringTerm T  + \sqrt{2DN} X) = \expEnv \left[ \mathbbm{1}_{\left\{R(NT) \geq c(N) T + \sqrt{2DN} X\right\}} \right]
\end{equation}
Defining (and then simplifying)
\begin{equation}\label{eq:tailTestFunction}
    \phi^{\mathrm{tail}}_N\left( X' \right) := \frac{C(N,T,X)}{C\left(N, T, X' \right)} \mathbbm{1}_{\left\{X' \geq X\right\}} =  \exp\left\{-\frac{N^{1/4}}{\sqrt{2D}}(X' - X)\right\}\mathbbm{1}_{\left\{X' \geq X\right\}}
\end{equation}
we can rewrite Eq. \ref{eq:tailphi} as 
\begin{equation}
\begin{aligned}\label{eq:TailAfterPhiSubstitution}
    \mathbb{P}^{\env}( R(NT) \geq \centeringTerm T  + \sqrt{2DN} X) &= \frac{1}{C(N,T,X)} \expEnv \left[ C\left(N, T, \frac{R^1( NT ) -  \centeringTerm T}{\sqrt{2DN}}\right) \phi^{\mathrm{tail}}_N\left( \frac{R^1( NT ) -  \centeringTerm T}{\sqrt{2DN}}\right) \right]\\
    &= \frac{1}{C(N, T, X)} \mathscr{U}_N(T, \phi),
\end{aligned}
\end{equation}
where we recall $\mathscr{U}_N(T, \phi)$ from Eq. \ref{eq:UdefExp}. 
Observe that as $N\to \infty$,
$$
\exp\left\{-\frac{N^{1/4}}{\sqrt{2D}}(X' - X)\right\}\mathbbm{1}_{\left\{X' \geq X\right\}} \approx \delta\left(\frac{N^{1/4}}{\sqrt{2D}}(X' - X) \right)\mathbbm{1}_{\left\{X' \geq X\right\}}.
$$
Note here that the convergence shown earlier was for fixed $\phi$ while we are now permitting $\phi$ to vary in $N$ and tend to a Dirac delta function. Thus, in light of the convergence of $\mathscr{U}_N(T, \phi)$ to the SHE, this implies that
\begin{align}
     \mathbb{P}^{\env}( R(NT) \geq \centeringTerm T  + \sqrt{2DN} X) &\approx \frac{1}{C(N,T,X)} \int_{X}^{\infty} \delta\left(-\frac{N^{1/4}}{\sqrt{2D}} (X' - X)\right) Z(T', X')dX'\\
    &= \frac{\sqrt{2D}}{N^{1/4}C(N,T,X)} Z(T, X).
\end{align}
Moving the prefactor to the left-hand side,
\begin{equation}
  \lim_{N\rightarrow\infty}  \frac{N^{1/4}C(N,T,X)}{\sqrt{2D}} \mathbb{P}^{\env}( R(NT) \geq \centeringTerm T  + \sqrt{2DN} X) = Z(T, X).
\end{equation}
Thus, the tail probability converges to the SHE with the same noise strength but a modified prefactor. 

\section{Conclusion}
We have demonstrated that under a moderate deviations scaling, there are universal KPZ fluctuations for a large class of RWRE models. We show that the strength of the noise of the KPZ equation is characterized by the variable $\extCoef$, which depends on the statistics of the underlying environment. Our results can be extended to characterize the distribution of the extreme value statistics of a system of $N$ diffusing particles as in \cite{hassAnomalousFluctuationsExtremes2023, hassFirstpassageTimeManyparticle2024, hassExtremeDiffusionMeasures2024a} (e.g., the first time a particle reaches a position $L$ or the position of the furthest particle at time $t$). By measuring these extreme value statistics, microscopic information of the environment can be studied via our derived prefactor, $\extCoef$. 
\end{widetext}
\bibliographystyle{unsrtnat} 
\bibliography{main}

\end{document}